\documentclass[10pt,superscriptaddress,twocolumn,amsmath,amssymb,aps,prb]{revtex4}
\usepackage{mathrsfs}
\usepackage{graphicx}% Include figure files
\usepackage{dcolumn}% Align table columns on decimal point
\usepackage{bm}% bold math
\usepackage{array}
\usepackage{booktabs}
\newcommand{\be}{\begin{equation}}
\newcommand{\ee}{\end{equation}}
\newcommand{\bea}{\begin{eqnarray}}
\newcommand{\eea}{\end{eqnarray}}

\begin{document}

\title{Topological superconductivity in Ni-based transition-metal trichalcogenides}
\author{Yinxiang Li}
\affiliation{Beijing National Laboratory for Condensed Matter Physics,
and Institute of Physics, Chinese Academy of Sciences, Beijing 100190, China}
\affiliation {Tin Ka-Ping College of Science, University of Shanghai for Science and Technology, Shanghai, 200093, China}
\author{Xianxin Wu}\email{xianxinwu@gmail.com}
\affiliation{Institute for Theoretical Physics and Astrophysics, Julius-Maximilians University of Wurzburg, Am Hubland, D-97074 Wurzburg, Germany}
\author{Yuhao Gu}
\affiliation{Beijing National Laboratory for Molecular Sciences, State Key Laboratory of Rare Earth
Materials Chemistry and Applications, Institute of Theoretical and Computational Chemistry,
College of Chemistry and Molecular Engineering, Peking University, 100871 Beijing, China}
\affiliation{Beijing National Laboratory for Condensed Matter Physics,
and Institute of Physics, Chinese Academy of Sciences, Beijing 100190, China}
\author{Congcong Le}
\affiliation{Kavli Institute of Theoretical Sciences, University of Chinese Academy of Sciences, Beijing, 100049, China}
\author{Shengshan Qin}
\affiliation{Kavli Institute of Theoretical Sciences, University of Chinese Academy of Sciences, Beijing, 100049, China}
\author{Ronny Thomale}
\affiliation{Institute for Theoretical Physics and Astrophysics, Julius-Maximilians University of Wurzburg, Am Hubland, D-97074 Wurzburg, Germany}
\author{Jiangping Hu}\email{jphu@iphy.ac.cn}
\affiliation{Beijing National Laboratory for Condensed Matter Physics,
and Institute of Physics, Chinese Academy of Sciences, Beijing 100190, China}
\affiliation{Kavli Institute of Theoretical Sciences, University of Chinese Academy of Sciences, Beijing, 100049, China}
\affiliation{Collaborative Innovation Center of Quantum Matter, Beijing 100049, China}
\date{\today}

\begin{abstract}
Based on a two-orbital honeycomb lattice model and by use of a random phase approximation analysis, we investigate the pairing symmetry of Ni-based transition-metal trichalcogenides. We find that an $I$-wave (A$_{2g}$) state and a chiral $d$-wave (E$_{g}$) state are dominant and nearly degenerate for typical electron and hole dopings.  Both states exhibit nontrivial topological properties, which manifest themselves by the chiral edge states for the $d+id$-wave state and dispersionless Andreev bound state at zero energy for the $I$-wave state. We thus show that Ni-based transition-metal trichalcogenides provide a promising platform to study exotic topological phenomena emerging from electronic correlation effects.
\end{abstract}

\pacs{74.20.Mn, 74.70.Dd, 74.20.Rp}

\maketitle
\section{Introduction}\label{sectioni}
Two-dimensional (2D) materials, such as graphene and transition-metal dichalcogenides \cite{Nov,Geim}, are under extensive contemporary study in condensed matter research. This is because many intriguing quantum phenomena have been realized in them, including spin-valley coupling\cite{Xu2014}, strong charge-spin correlation \cite{Kim}, 2D magnetism \cite{Kim2,Wildes,Gong,Huang} and even superconductivity\cite{Lu,Xi,Saito,Costanzo,Wang}. Among them, ternary transition-metal phosphorus trichalcogenide (TMPT) compounds APX$_{3}$ (A=3d transition metals; X=chalcogens) have attracted enormous attention due to antiferromagnetic (AF) ordering as a hint for significant electronic correlations. For bulk materials, it is experimentally found that they can exhibit diverse AF structures, such as zigzag AF and stripy AF\cite{Chittari,Sivadas,Lee,Wildes}. They can be further exfoliated into few atomic layers \cite{Kuo}, rendering them extremely suitable for studying AF ordering in the quasi-2D limit. Furthermore, by suppressing AF order with external pressure, superconductivity emerges in iron-based TMPT compounds such as FePSe$_{3}$, with the highest $T_c$ found at about 5.5 K \cite{Wang}, bearing resemblance to high T$_c$ cuprates and iron based superconductors. All this accumulated evidence strongly suggests that electronic correlations may be of great importance to the family of TMPTs.

The crystal structure of the TMPT family APX$_{3}$ consists of edge shared AX$_{6}$ octahedral complexes and P2 dimers. Transition metal atoms are arranged in a hexagonal lattice, as shown in Figs.\ref{fig1}(a) and (b). In the octahedral crystal field, five 3d orbitals of transition-metal atoms split into high-energy e$_{g}$ orbitals and low-energy t$_{2g}$ orbitals. For FePX$_{3}$ with Fe$^{2+}$ ions (d$^6$), it is an ideal system to study the high-to-low spin-state transition by pressure\cite{Wang}. For the case of NiPX$_{3}$ with Ni$^{2+}$ $d^{8}$ filling configuration, t$_{2g}$ bands are fully occupied while e$_{g}$ bands are half filled and dominate the spectral weight near the Fermi level. Therefore, the low energy physics can be described by a two orbital model on the honeycomb lattice, where Dirac cones at K are expected. Ni-based phosphorus trichalcogenides, however, can host additional Dirac cones around the midpoint $\frac{K}{2}$ of $\Gamma K$ near Fermi level\cite{YHGu}. The 3d orbital nature renders them ideal candidates for strongly correlated Dirac electron system. According to density functional theory (DFT) calculation and experimental measurements, the ground state of Ni-based phosphorus trichalcogenides has been found to be zigzag AF insulator \cite{Flem,Chittari,YHGu}. Theoretical calculations suggests that charge doping can suppress magnetic order, and superconductivity can eventually be achieved\cite{YHGu}.

In this article, we investigate the pairing symmetry of Ni-based transition-metal trichalcogenide superconductors near half filling. Based on a two-orbital Hubbard model on the honeycomb lattice and through the use of a random phase approximation (RPA) analysis, we find that an $I$-wave and a chiral $d$-wave superconducting state are the dominant instabilities, and nearly degenerate in terms of pairing propensity for typical electron and hole doping. This is because both instabilities are promoted by the intra Fermi surface nesting (between $\alpha$ FS) and the inter Fermi surface nesting (between $\alpha$ and $\beta$ FS). By further resolving their real-space pairing structure, we find that the $I$-wave ($A_{2g}$) state is mainly attributed to NN and NNN pairing, while the chiral $d$-wave E$_{g}$ state is broadly extended in real space. Both superconducting orders involve strong inter orbital pairing and exhibit nontrivial topological properties. The chiral $d$-wave state is characterized by a nontrivial Chern number, and as such chiral edge modes. In turn, the nodal $A_{2g}$ state features a non-trivial one dimensional topological invariant, and flat bands at zero energy can appear on the edges. Due to their distinctive physical properties, a variety of experimental measurements can be used to unambiguously distinguish these two pairing states.

The paper is organized as follows. In Sec. \ref{sectionii}, we present the two-orbital tight-binding model based on e$_{g}$ orbitals (d$_{xz}$ and d$_{yz}$ orbitals) to represent the single-particle description of the Ni-based transition-metal trichalcogenide. Furthermore, the crystal structure and electronic band structure is discussed. In Sec. \ref{sectioniii}, we explain the formalism of the random phase approximation (RPA) approach we employ to predict the pairing symmetries within the scope of multi-orbital Coulomb interactions. In Sec. \ref{sectioniv}, we calculate the spin susceptibility and pairing symmetry as a function of electron and hole doping starting from half filling. We also analyze the real-space pairing of the obtained pairing states, and calculate the edge states originating from their nontrivial topological properties. Finally, in Sec. \ref{sectionv} we discuss about the experimental realization of superconductivity conclude that transition-metal trichalcogenide provide a promising platform for topological superconductivity.

\section{Nickel phosphorous trichalcogenides}\label{sectionii}
\begin{figure}
\centerline{\includegraphics[width=0.5\textwidth]{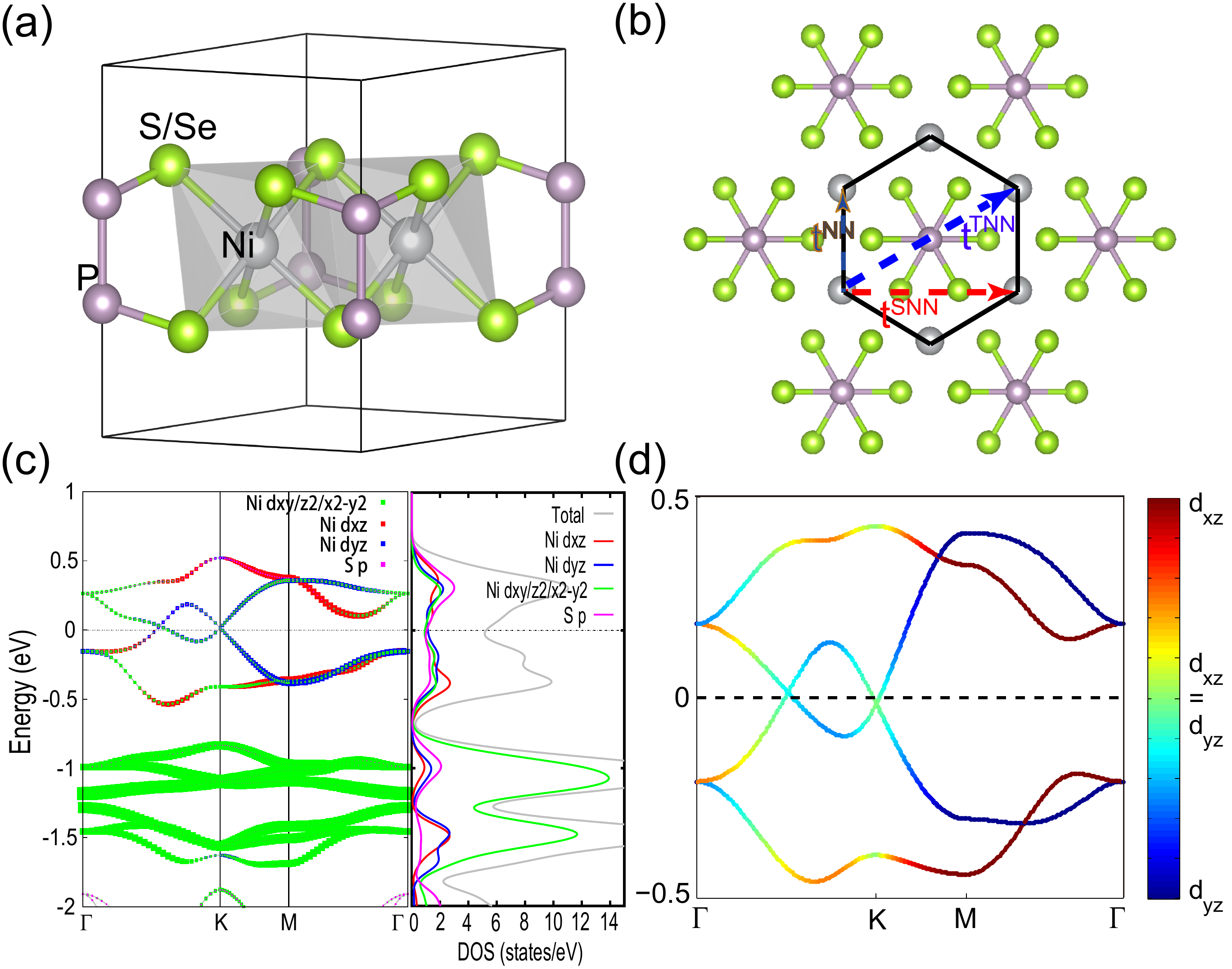}}
\caption{(color online) (a) The crystal structures of the monolayer NiPX$_{3}$ (X=S, Se) (space group P-31m). (b) The top view of the monolayer NiPX$_{3}$. (c) Electronic band structure and density of states (DOS) for NiPS$_3$. The orbital characters of bands are represented by different colors. (d) Band dispersion of tight-binding model based on d$_{xz}$ and d$_{yz}$ orbitals of Ni atoms.}
\label{fig1}
\end{figure}

The compounds nickel phosphorous trichalcogenides NiPX$_{3}$ (X=S,Se) crystallizes in a layered hexagonal structure and each layer consists of edge shared MX$_6$ octahedral complexes and P$_2$ dimers. As shown in Fig. \ref{fig1}(a) and (b), the cation Ni is surrounded by six anions S/Se and the anions entity $P_{2}X^{-4}_{6}$ is located at the center of honeycomb lattice. As they can be easily exfoliated to monolayers, we focus on NiPX$_{3}$ monolayers in the following. The electronic band structure and Density of states for monolayer NiPS$_{3}$ are displayed in Fig. \ref{fig1}(c). It is evident that the $e_g$ and $t_{2g}$ bands is separated by a gap due to the crystal field. The former bands with high energies appear near the Fermi level while the latter locate around $1.4$ eV below the Fermi level. The $e_{g}$ bands ($d_{xz}$, $d_{yz}$ orbitals) are half filled and contribute dominantly to the Fermi surface. Moreover, their bandwidth is quite narrow only about 1.1 eV. The Dirac cone appears at K(K') is similar to that in graphene but additional Dirac cones appear around $\frac{K}{2}$ ($\frac{K'}{2}$), which is protected by the mirror symmetry along $GK$ line. The basic electronic structure can be modelled by a two-orbital tight-binding model on a honeycomb lattice. The corresponding Hamiltonian reads,
\begin{equation}
\emph{H}_{0}=\sum_{\bm{k}\alpha\beta}\sum_{l_1l_2\sigma}h^{\alpha\beta}_{l_1l_2}(k)c^{\dag}_{\bm{k}\alpha l_1\sigma}c_{\bm{k}\beta l_2\sigma}=\sum_{\bm{k}}\psi^\dag_{\bm{k}\sigma}h(\bm{k})\psi_{\bm{k}\sigma},
\end{equation}
with $\psi^\dag_{\bm{k}\sigma}=(c^{\dag}_{\bm{k}A1\sigma},c^{\dag}_{\bm{k}A2\sigma},c^{\dag}_{\bm{k}B1\sigma},c^{\dag}_{\bm{k}B2\sigma})$.
Here $\alpha$,$\beta$ are the sublattice indices (A,B) and $l_i$ ($l_i=1,2$) is the orbital index ($d_{xz/yz}$). $c^{\dag}_{\alpha\mu\sigma}$ creates a spin $\sigma$ electron with momentum $\bm{k}$ in $\mu$ orbital on $\alpha$ sublattice. The matrix elements of $h^{\alpha\beta}_{\mu\nu}(k)$ are provided in the Appendix. According to our calculations\cite{YHGu}, we interestingly find the third nearest neighbor (TNN) hopping is much larger than the NN and second nearest neighbor (SNN) hopping parameters. From the calculation of GGA+$U$\cite{YHGu}, the NiPS$_{3}$ favors the zigzag antiferromagnetic state, where the magnetic moments of Ni cations connected by the TNN bonds is antiparallel. We show the orbital resolved band dispersion from the tight-binding model in Fig. \ref{fig1}(d). Orbital mixture can be found along $\Gamma K$ and $KM$ but not $\Gamma M$, due to the presence of two-fold rotational symmetry along $\Gamma M$. The strongest orbital mixture near the Fermi level occurs around the Dirac points ($K$ and $\frac{K}{2}$).

\section{random phase approximation}\label{sectioniii}
In this section, we explain the formalism of the multiorbital RPA approach\cite{Bickers,Kemper,Graser,Xxwu1,Xxwu2,Li,Sante}. The adopted onsite Coulomb interaction terms are,
\begin{equation}
\begin{aligned}
H_{int}=&U\sum_{i,\alpha}n_{i\alpha\uparrow}n_{i\alpha\downarrow}+U'\sum_{i,\alpha<\beta}n_{i\alpha}n_{i\beta}\\
         &+J_{H}\sum_{i,\alpha<\beta,\sigma\sigma^{'}}c^{\dagger}_{i\alpha\sigma}c^{\dagger}_{i\beta\sigma^{'}}c_{i\alpha\sigma^{'}}c_{i\beta\sigma}\\
         &+J'\sum_{i,\alpha\neq\beta}c^{\dagger}_{i\alpha\uparrow}c^{\dagger}_{i\alpha\downarrow}c_{i\beta\downarrow}c_{i\beta\uparrow},
\end{aligned}
\label{hh}
\end{equation}
where $n_{i,\alpha}=n_{i,\alpha,\uparrow}+n_{i,\alpha,\downarrow}$. $U$, $U'$, $J$ and $J'$ represent the intra- and inter-orbital repulsion, the Hund's rule and pair-hopping terms. In the following calculations, we use Kanamori relations $U=U'+2J$ and $J=J'$ as requried by the lattice symmetry.

The multi-orbital susceptibility is defined as,
\begin{eqnarray}
\chi_{l_1l_2l_3l_4}(\bm{q},\tau)&=&\frac{1}{N}\sum_{\bm{k}\bm{k}'}\langle T_{\tau} c^{\dag}_{l_3\sigma}(\bm{k}+\bm{q},\tau)c_{l_4\sigma}(\bm{k},\tau)\nonumber\\
&&c^{\dag}_{l_2\sigma'}(\bm{k}'-\bm{q},0)c_{l_1\sigma'}(\bm{k}',0) \rangle .
\end{eqnarray}
In momentum-frequency space, the multi-orbital bare susceptibility is given by
\begin{equation}
\begin{aligned}
&\chi^{0}_{l_{1}l_{2}l_{3}l_{4}}(\bm{q},i\omega_{n})=-\frac{1}{N}\sum_{\bm{k}\mu\nu}a^{l_{4}}_{\mu}(\bm{k})a^{l_{2}*}_{\mu}(\bm{k})a^{l_{1}}_{\nu}(\bm{k}+\bm{q})\times\\
&a^{l_{3}*}_{\nu}(\bm{k}+\bm{q})\frac{n_{F}(E_{\mu}(\bm{k}))-n_{F}(E_{\nu}(\bm{k}+\bm{q}))}{i\omega_{n}+E_{\mu}(\bm{k})-E_{\nu}(\bm{k}+\bm{q})},
\end{aligned}
\end{equation}
where $\mu$ and $\nu$ are the band indices, $n_{F}$ is the usual Fermi distribution, $l_{i}$ $(i=1,2,3,4)$ are the orbital indices, $a^{l_{i}}_{\mu}(k)$ is the $l_{i}$ orbital component of the eigenvector for band $\mu$ resulting from the diagonalization of the tight-binding Hamiltonian $H_{0}$ and $E_{\mu}(\bm{k})$ is the corresponding eigenvalue. With interactions, the RPA spin and charge susceptibilities are given by
\begin{equation}
\begin{aligned}
\chi^{RPA}_{s}(\bm{q})=\chi^{0}(\bm{q})[1-\bar{U}^{s}\chi^{0}(\bm{q})]^{-1},\\
\chi^{RPA}_{c}(\bm{q})=\chi^{0}(\bm{q})[1+\bar{U}^{c}\chi^{0}(\bm{q})]^{-1},
\end{aligned}
\end{equation}
where $\bar{U}^{s}$ ($\bar{U}^{c}$) is the spin (charge) interaction matrix
\begin{displaymath}
     \bar{U}^{s/c}=
     \left(\begin{array}{cc}
           \bar{U}^{s/c}_{A}&0\\
           0&\bar{U}^{s/c}_{B}
           \end{array}
     \right),
\end{displaymath}

$$ \bar{U}^{s}_{A/B,l_{1}l_{2}l_{3}l_{4}}=\left\{
\begin{array}{rcl}
\begin{aligned}
&U                & {l_{1}=l_{2}=l_{3}=l_{4}},\\
&U'            & {l_{1}=l_{3}\neq l_{2}=l_{4}},\\
&J                & {l_{1}=l_{2}\neq l_{3}=l_{4}},\\
&J'            & {l_{1}=l_{4}\neq l_{3}=l_{2}},
\end{aligned}
\end{array} \right. $$

$$ \bar{U}^{c}_{A/B,l_{1}l_{2}l_{3}l_{4}}=\left\{
\begin{array}{rcl}
\begin{aligned}
&U                  & {l_{1}=l_{2}=l_{3}=l_{4}},\\
&-U'+2J          & {l_{1}=l_{3}\neq l_{2}=l_{4}},\\
&2U'-J           & {l_{1}=l_{2}\neq l_{3}=l_{4}},\\
&J'              & {l_{1}=l_{4}\neq l_{3}=l_{2}},
\end{aligned}
\end{array} \right. $$
Within RPA approximation, the effective Cooper scattering interaction is,
\begin{equation}
\begin{aligned}
\Gamma_{ij}(\bm{k},\bm{k}')=&\sum_{l_{1}l_{2}l_{3}l_{4}}a^{l_{2},\ast}_{\emph{v}_{i}}(\bm{k})a^{l_{3},\ast}_{\emph{v}_{i}}(-\bm{k})\\
&\times\emph{Re}\bigg[\Gamma_{l_{1}l_{2}l_{3}l_{4}}(\bm{k},\bm{k}',\omega=0)\bigg]a^{l_{1}}_{\emph{v}_{j}}(\bm{k}')a^{l_{4}}_{\emph{v}_{j}}(-\bm{k}'),
\end{aligned}
\end{equation}
where the momenta $\bm{k}$ and $\bm{k}'$ is restricted to different FSs with $\bm{k}\in C_{i}$ and $\bm{k}'\in C_{j}$. The  orbital vertex function $\Gamma_{l_{1}l_{2}l_{3}l_{4}}$ in spin singlet and triplet channels \cite{Kemper,Xxwu2,Sante} are
\begin{equation}
\begin{aligned}
\Gamma^{S}_{l_{1}l_{2}l_{3}l_{4}}(\bm{k},\bm{k}',\omega)=&\bigg[\frac{3}{2}\bar{U}^{s}\chi^{RPA}_{s}(\bm{k}-\bm{k}',\omega)\bar{U}^{s}+\frac{1}{2}\bar{U}^{s}\\
-&\frac{1}{2}\bar{U}^{c}\chi^{RPA}_{c}(\bm{k}-\bm{k}',\omega)\bar{U}^{c}+\frac{1}{2}\bar{U}^{c}\bigg]_{l_{1}l_{2}l_{3}l_{4}},\\
\Gamma^{T}_{l_{1}l_{2}l_{3}l_{4}}(\bm{k},\bm{k}',\omega)=&\bigg[-\frac{1}{2}\bar{U}^{s}\chi^{RPA}_{s}(\bm{k}-\bm{k}',\omega)\bar{U}^{s}+\frac{1}{2}\bar{U}^{s}\\
-&\frac{1}{2}\bar{U}^{c}\chi^{RPA}_{c}(\bm{k}-\bm{k}',\omega)\bar{U}^{c}+\frac{1}{2}\bar{U}^{c}\bigg]_{l_{1}l_{2}l_{3}l_{4}},
\end{aligned}
\end{equation}
where $\chi^{RPA}_{s}$ and $\chi^{RPA}_{c}$ are the RPA spin and charge susceptibility, respectively. The pairing strength functional for a specific pairing state is given by,
\begin{equation}
\begin{aligned}
\lambda\big[\emph{g}(\bm{k})\big]=-\frac{\sum_{ij}\oint_{C_{i}}\frac{d\bm{k}_{\|}}{\emph{v}_{\emph{F}}(\bm{k})}\oint_{C_{j}}\frac{d\bm{k}'_{\|}}{\emph{v}_{\emph{F}}(\bm{k}')}\emph{g}(\bm{k})\Gamma_{ij}(\bm{k},\bm{k}')\emph{g}(\bm{k}')}{(2\pi)^{2}\sum_{i}\oint_{C_{i}}\frac{d\bm{k}_{\|}}{\emph{v}_{\emph{F}}(\bm{k})}\big[\emph{g}(\bm{k})\big]^{2}},
\end{aligned}
\end{equation}
where $v_{F}(\bm{k})=|\nabla_{k}E_{i}(\bm{k})|$ is the Fermi velocity on a given Fermi surface sheet $C_{i}$. The pairing vertex function in spin singlet and triplet channels are symmetric and antisymmetric parts of the interaction, that is, $\Gamma^{S/T}_{ij}(\bm{k},\bm{k}')=\frac{1}{2}[\Gamma_{ij}(\bm{k},\bm{k}')\pm \Gamma_{ij}(\bm{k},-\bm{k}')]$.

\begin{figure}
\centerline{\includegraphics[width=0.5\textwidth]{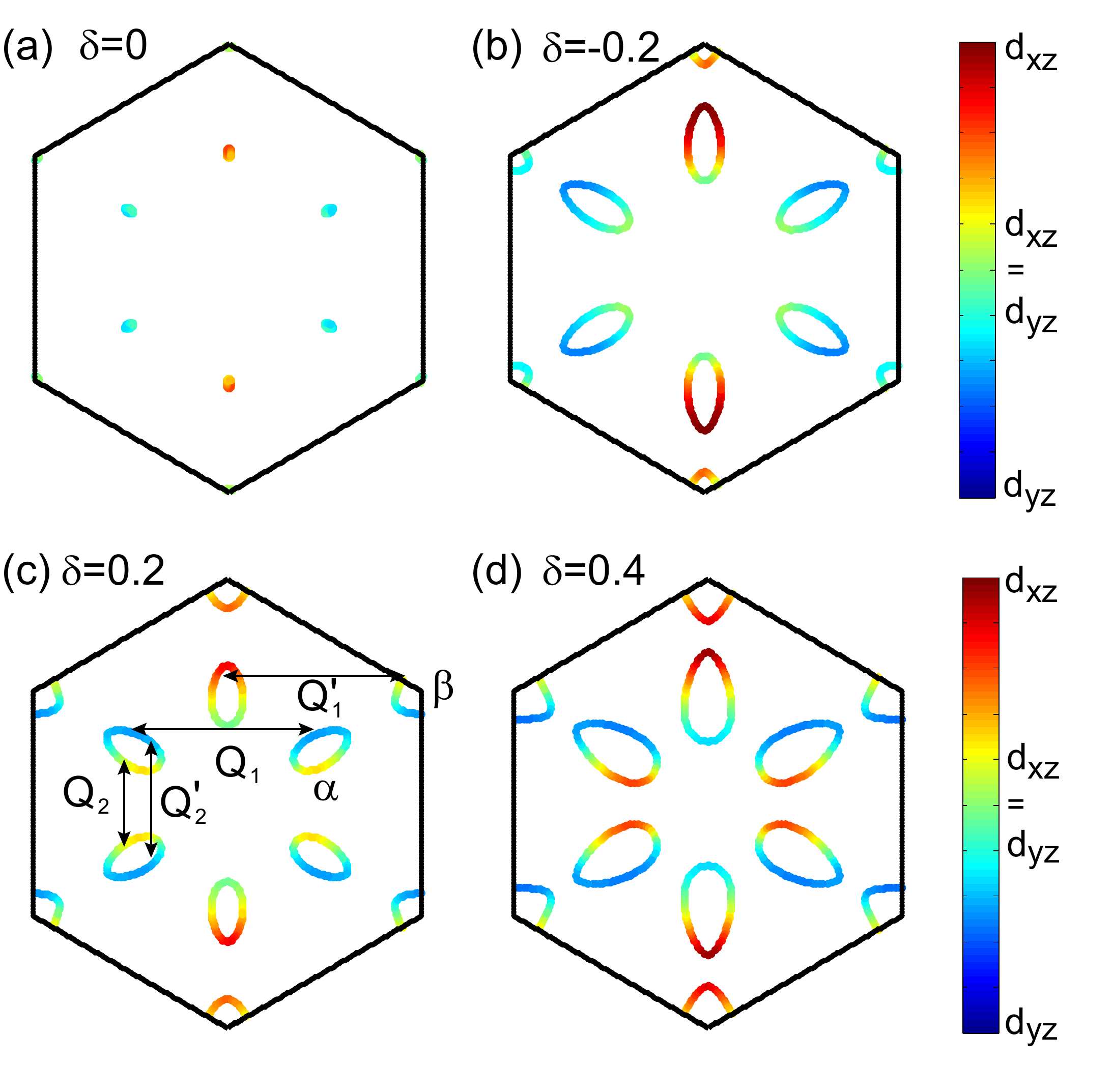}}
\caption{(color online) Orbital resolved Fermi surface with different electron filling. (a) $\delta=0$, (b) $\delta=-0.2$, (c) $\delta=0.2$ and (d) $\delta=0.4$. The Fermi surfaces around $K$ and $\frac{K}{2}$ are represented by $\alpha$ and $\beta$. }
\label{fig2_FS}
\end{figure}

\begin{figure}
\centerline{\includegraphics[width=0.5\textwidth]{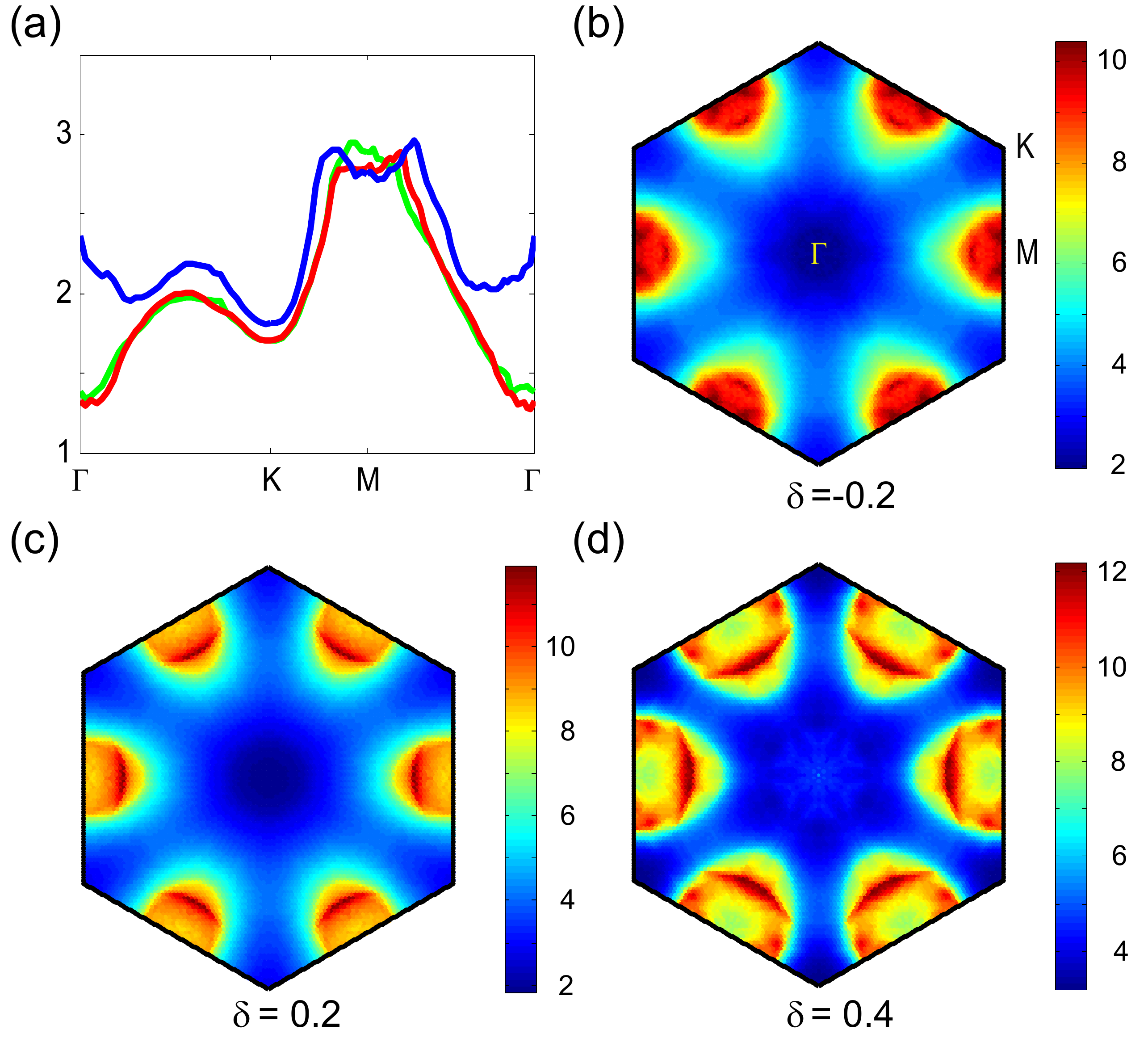}}
\caption{(color online) Susceptibilities for different doping. (a) bare susceptibilities along high-symmetry lines for $\delta=$ -0.2 (green line), 0.2 (red line) and 0.4 (blue line). The corresponding RPA spin susceptibilities in Brillouin zone are shown in (b), (c) and (d) with $U=0.4$ and $J/U=0.2$.}
\label{fig2_sus}
\end{figure}

\begin{figure}
\centerline{\includegraphics[width=0.5\textwidth]{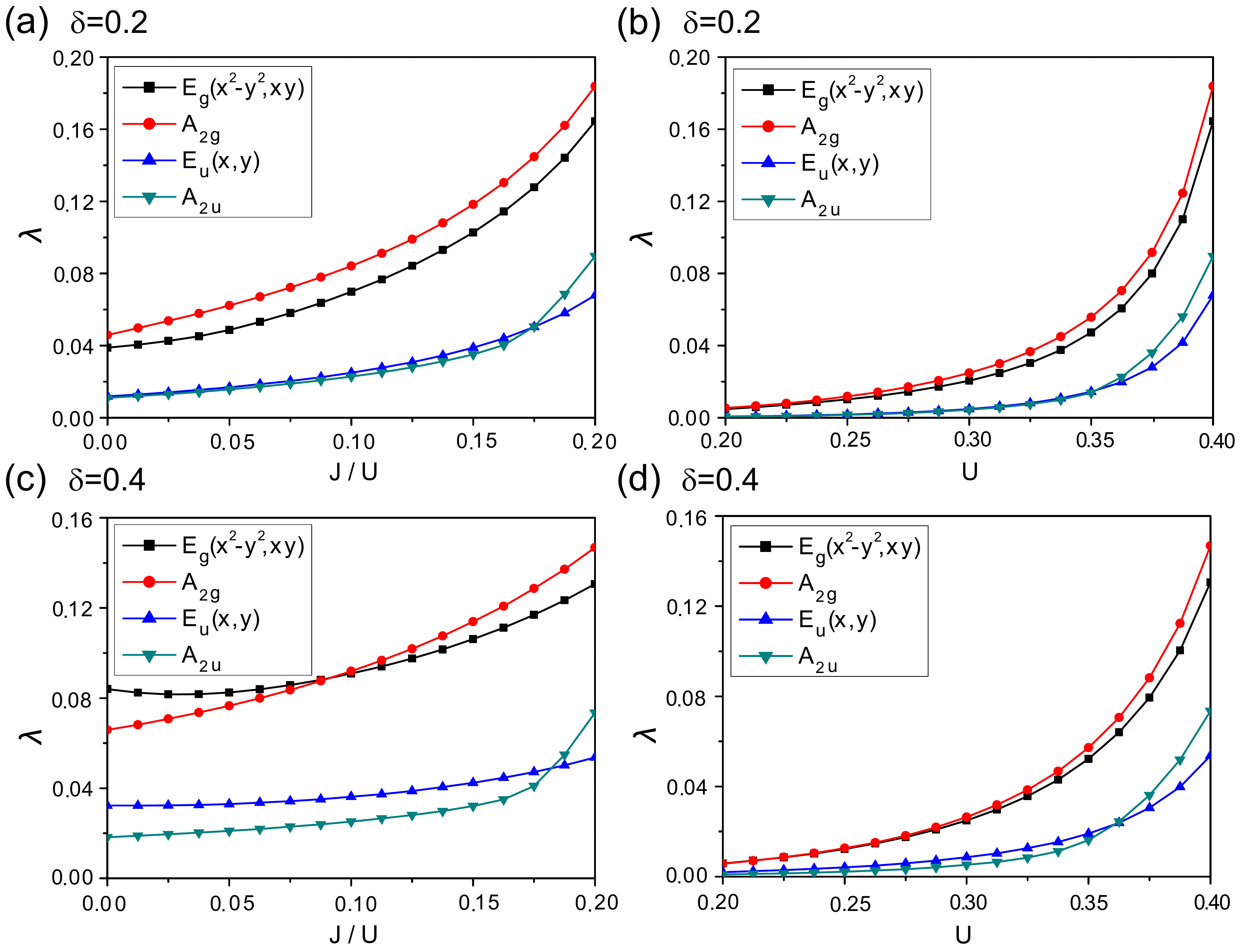}}
\caption{(color online) Leading pairing strengths in spin singlet and triplet channels for superconducting states with different doping near the half filling. The pairing strengths as a function of Hund's rule coupling parameter J/U with 0.2 and 0.4 electron doping at U=0.4 are plotted in (a) and (c). The pairing strengths as a function of on-site intraorbital Coulomb interaction U with J/U=0.2 with 0.2 and 0.4 electron doping are plotted in (b) and (d).}
\label{lamda_ed}
\end{figure}

\section{results and analysis}\label{sectioniv}
According to Ref.\onlinecite{YHGu}, superconductivity may emerge after the antiferromagnetic states is suppressed by electron or hole doping. Here, based on weak coupling appoach, we investigate the pairing symmetries for the doped system. In the following, we mainly focus on three typical doping levels $\delta=-0.2, 0.2, 0.4$ relative to the half filled bands. $\delta>0$ ($\delta<0$) represents the electron (hole) doping. Fig. \ref{fig2_FS} shows the Fermi surfaces for half-filled case and the above doped cases. At half filling, Dirac cones at K and $K'$ contributes two small electron pockets while the Dirac cones around $\frac{K}{2}$ and $\frac{K'}{2}$ contribute six small hole pockets, whose area is equal to that of electron pockets as required by charge neutrality. With hole doping, the Fermi surfaces around $K$ shrink to the eletron-hole Lifshitz transition point and then become hole pockets while $\alpha$ Fermi surfaces enlarge and become elliptical, with the long axis along $\Gamma K$. Two of $\alpha$ Fermi surfaces are mainly attributed to $d_{xz}$ orbital while the others are mainly attributed to $d_{yz}$ orbitals. The parts of $\alpha$ Fermi surfaces close to $\Gamma$ point exhibit strong orbital mixture while those away from $\Gamma$ point show weaker orbital mixture. For 0.2 electron doped case, as shown in Fig.\ref{fig2_FS}(c), all Fermi surfaces are electron type and the orbital distribution on the Fermi surfaces are similar. With further electron doping, $\delta=0.4$, all $\alpha$ Fermi surfaces enlarge and show a stronger orbital mixture.

% In multi-orbital RPA scheme, on-site repulsive Hubbard interactions enhance the spin susceptibility and nearest-neighbor site repulsive interactions motivate the enhancement of charge susceptibility.
%The divergence of spin (charge) susceptibility implies that the system comes into the state of spin density wave (charge density wave). In our calculation, the spin susceptibility diverges nearly at $U_{c}=0.45$ which means the invalidity of RPA treatment.
%  at $U=0.4$, $J/U=0.2$, $J=J^{'}$ and $U^{'}=U-2J$. the RPA susceptibility in RPA calculation $\chi^{RPA}_{s}$ in Fig. \ref{fig2_sus} for the three typical doping levels at $U=0.4$, $J/U=0.2$, $J=J^{'}$ and $U^{'}=U-2J$.

In order to study the pairing symmetries, we first show the bare susceptibility $\chi_{0}$ along high-symmetry lines in Fig. \ref{fig2_sus}(a) for the three typical doping levels. For $\delta=0.2$ (red line),
there is a prominent plateau around M and a broad peak around $\frac{K}{2}$. The former one is attributed to inter pocket nesting between $\alpha$ and $\beta$ and intra pocket nesting
between next NN of $\alpha$ Fermi surfaces. The corresponding nesting vectors $Q_1$ and $Q'_1$ are displayed in Fig.\ref{fig2_FS}(c). While the latter peak is mainly contributed by the intra pocket
nesting between the NN of $\alpha$ Fermi surfaces ($Q_2$ and $Q'_2$ ), as shown in Fig.\ref{fig2_FS}(c). With further increasing electron doping, the slight increase near M point results a double-peak structure and the broad peak enhances with its center
shifting toward K. In addition, a slight enhancement appears near $\Gamma$ point. The increase of the ratio $|\chi(K/2)|/|\chi(M)|$ is ascribed to the enlargement of $\alpha$ and $\beta$ pockets.
For the 0.2 hole doped case, the basic features are similar to those of electron doped case except that the double peaks merge into a single broad peak around M. We further calculate the RPA spin
susceptibility with $U=0.4$ eV, $J/U=0.2$ and the obtained results are displayed in Fig. \ref{fig2_sus}(b), (c) and (d). There are great enhancements for the peak structures mentioned above.
All peaks in the susceptibility are far away from $\Gamma$ point which implies the intrinsic antiferromagnetic fluctuations in the system, which is consistent the AF ordering in DFT calculations.

\begin{figure}
\centerline{\includegraphics[width=0.5\textwidth]{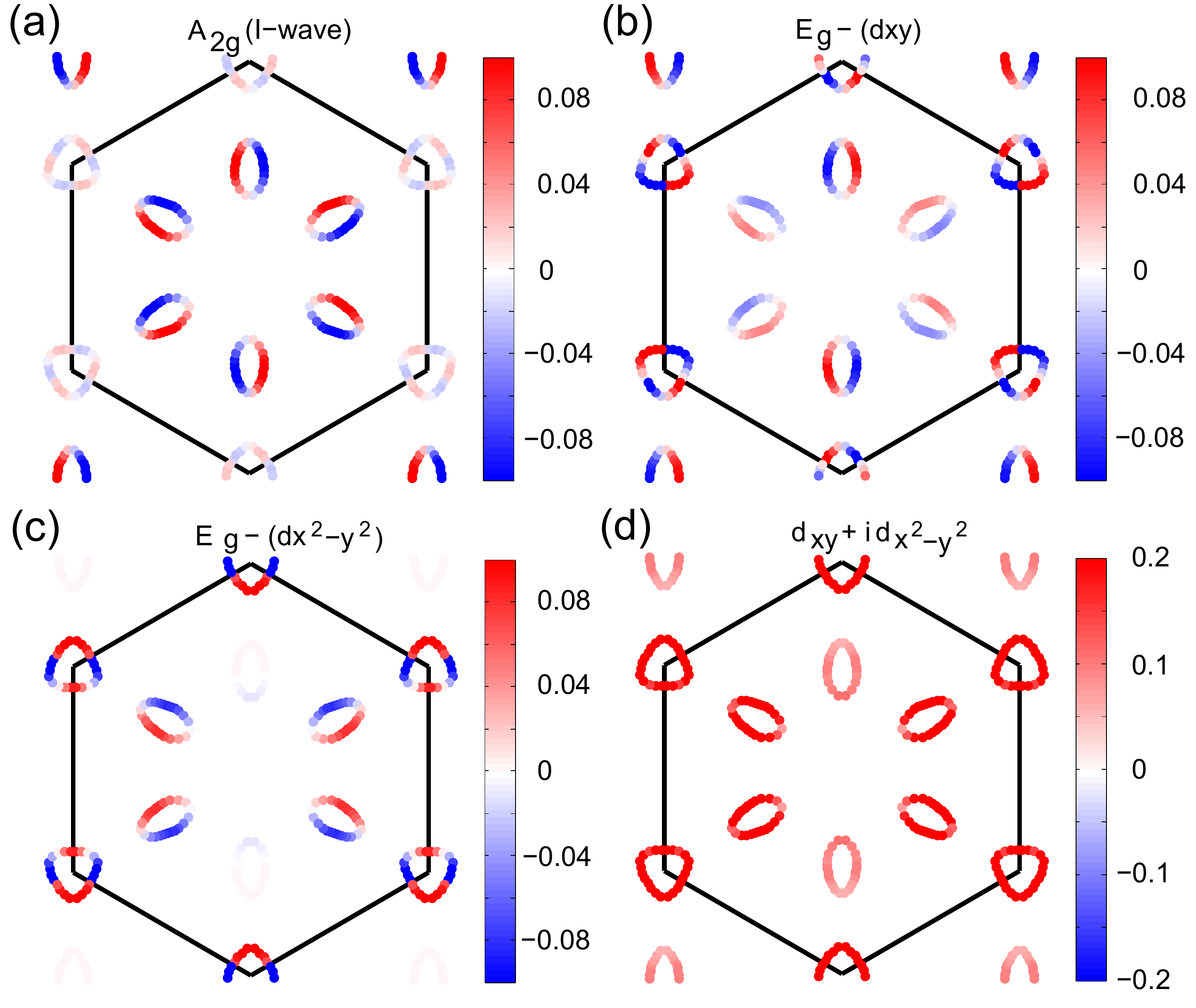}}
\caption{(color online) The gap functions of two leading singlet pairings with U=0.4 and J/U=0.2 at 0.2 electron doping. (a) gap function of $A_{2g}$ (I-wave) irreducible representation. (b), (c) gap functions of $d_{xy}$ and $d_{x^{2}-y^{2}}$ wave in $E_{g}$ irreducible representation. (d) gap function of $d_{xy}$+$id_{x^{2}-y^{2}}$.}
\label{pair_occ42}
\end{figure}

The pairing states can be classified according to the irreducible representations of $D_{3d}$ point group for NiPS$_{3}$. First we discuss about the case with electron doping. We consider a fixed $U=0.4$ and Fig.\ref{lamda_ed}(a) and (c) show the pairing strength eigenvalues for the leading eigenvalues in singlet and triplet channels as a function of $J/U$ for $\delta=0.2$ and $0.4$. In both cases, the pairing strengths of $A_{2g}$ and $E_g$ pairing states are close but much stronger than those of triplet parings, which is consistent with the intrinsic antiferromagnetic fluctuations from the susceptibility analysis. For $\delta=0.2$, $A_{2g}$ is slightly favored than $E_g$ state for $J/U<0.2$. However, for $\delta=0.4$, the dominant pairing state is $E_g$ state for $J/U<0.08$ and $A_{2g}$ state will win for $J/U>0.08$. We further plot the pairing strengths as a function of $U$ with a fixed $J/U=0.2$ in Fig.\ref{lamda_ed}(b) and (d). The pairing strengths increase rapidly with increasing $U$ and the dominant pairing states are still $A_{2g}$ and $E_g$.

The gap functions for the dominant pairing states are shown in Fig.\ref{pair_occ42} for $\delta=0.2$. A$_{2g}$ state is invariant under $C_{3z}$ and $S_{6z}$ operations and changes sign under $M_{yz}$ and $C_{2x}$ operations. The corresponding pairing has nodes along $k_x=0$ , $k_y=0$, $k_x=\pm\sqrt{3}k_y$, $k_x=\pm\frac{1}{\sqrt{3}}k_y$ lines and can be described by a function $xy(x^{2}-3y^{2})(3x^{2}-y^{2})$, which is an $I$-wave pairing state. The gap function on $\alpha$ FS is much large than that on $\beta$ FS. For the two-fold degenerate $E_{g}$ $d_{xy}$ and $d_{x^2-y^2}$ states are displayed in Fig.\ref{pair_occ42}(b) and (c). For the $d_{xy}$ state, the gap function on each $\alpha$ pocket has a sign change and the resulting nodes are not along high-symmetry lines, different from that of $A_{2g}$ state. Moreover, in contrast to $A_{2g}$ state, the gap function on $\beta$ FS is comparable to that of $\alpha$ FS. For the $d_{x^2-y^2}$ state, two $\alpha$ pockets on $y$ axis have a small gap size compared with others. Both $A_{2g}$ and $E_g$ states satisfy the condition that the superconducting order connected by the nesting vectors ($Q_1$, $Q'_1$, $Q_2$ and $Q'_2$) has a sign change. In $A_{2g}$ state, intra pocket scattering for $\alpha$ FS plays the dominant role. While, in $E_g$ state, both intra pocket and inter pocket scattering are important. Furthermore, the two-fold degenerate $E_g$ states tend to form the $d+id$ state in order to maximize the condensation energy, and the gap function is shown in Fig.\ref{pair_occ42}(d). This is a recurrent theme for other instances of predicted chiral $d$-wave superconductivity for hexagonal lattices~\cite{Graphene3RG,GrapheneFRG,GrapheneFRG2,CobaltatesFRG,PhysRevB.89.020509}.

\begin{figure}
\centerline{\includegraphics[width=0.5\textwidth]{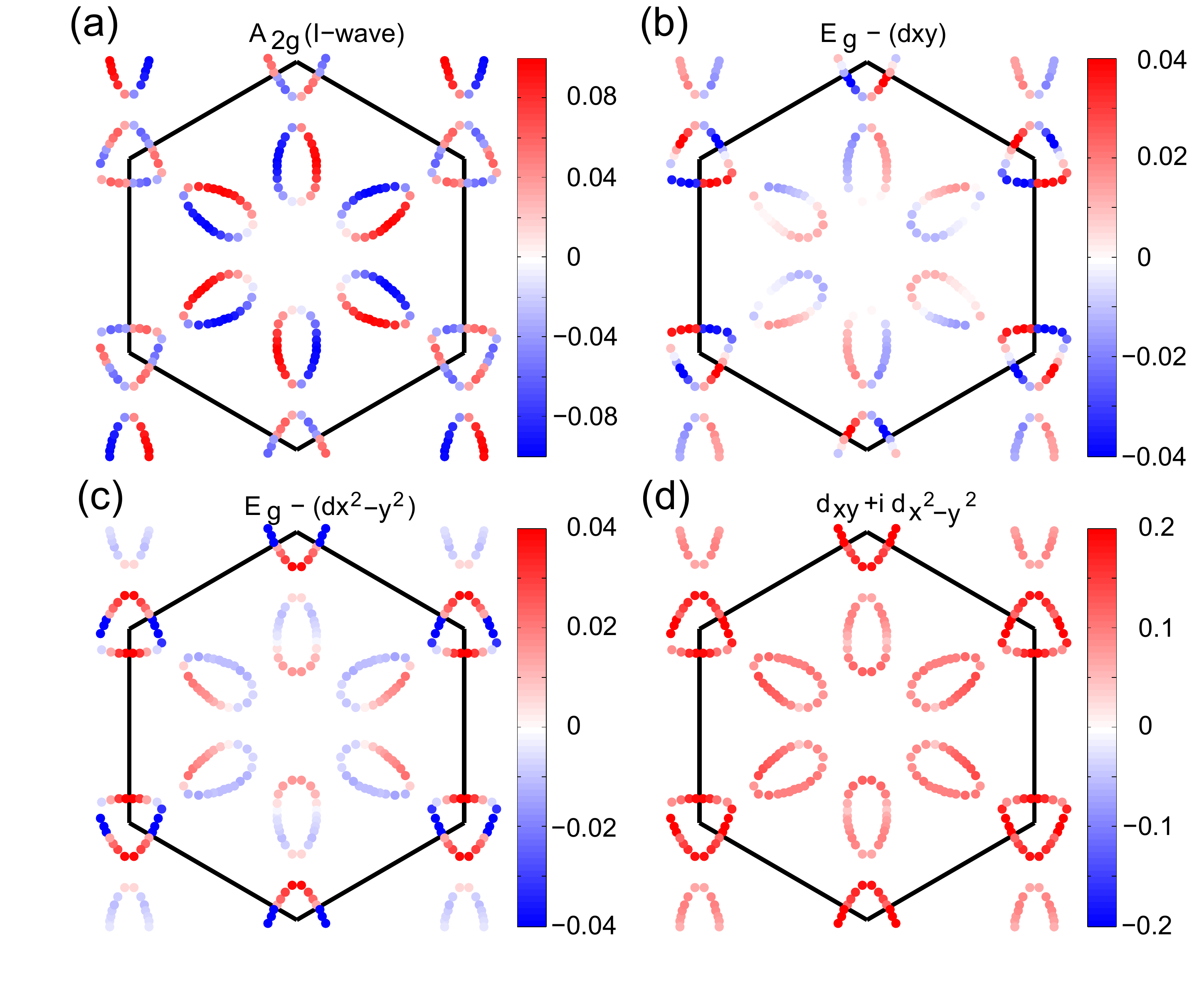}}
\caption{(color online) The gap functions of two leading singlet pairings with U=0.4 and J/U=0.2 at 0.4 electron doping. (a) gap function of $A_{2g}$ (I-wave) irreducible representation. (b), (c) gap functions of $d_{xy}$ and $d_{x^{2}-y^{2}}$ wave in $E_{g}$ irreducible representation. (d) gap function of $d_{xy}$+$id_{x^{2}-y^{2}}$.}
\label{pair_occ44}
\end{figure}

For $\delta=0.4$, the dominant gap functions are shown in Fig.\ref{pair_occ44}. There is a relative enhancement for gap functions on $\beta$ FS in $A_{2g}$ due to the enhancement of $Q'_1$ nesting. The inter pocket $Q'_1$ nesting enhancement also promote $E_g$ state, which explains the stronger pairing strength for $E_g$ for $J/U <0.08$. The corresponding gap functions are shown in Fig.\ref{pair_occ44}(b), (c) and (d).

\begin{figure}
\centerline{\includegraphics[width=0.5\textwidth]{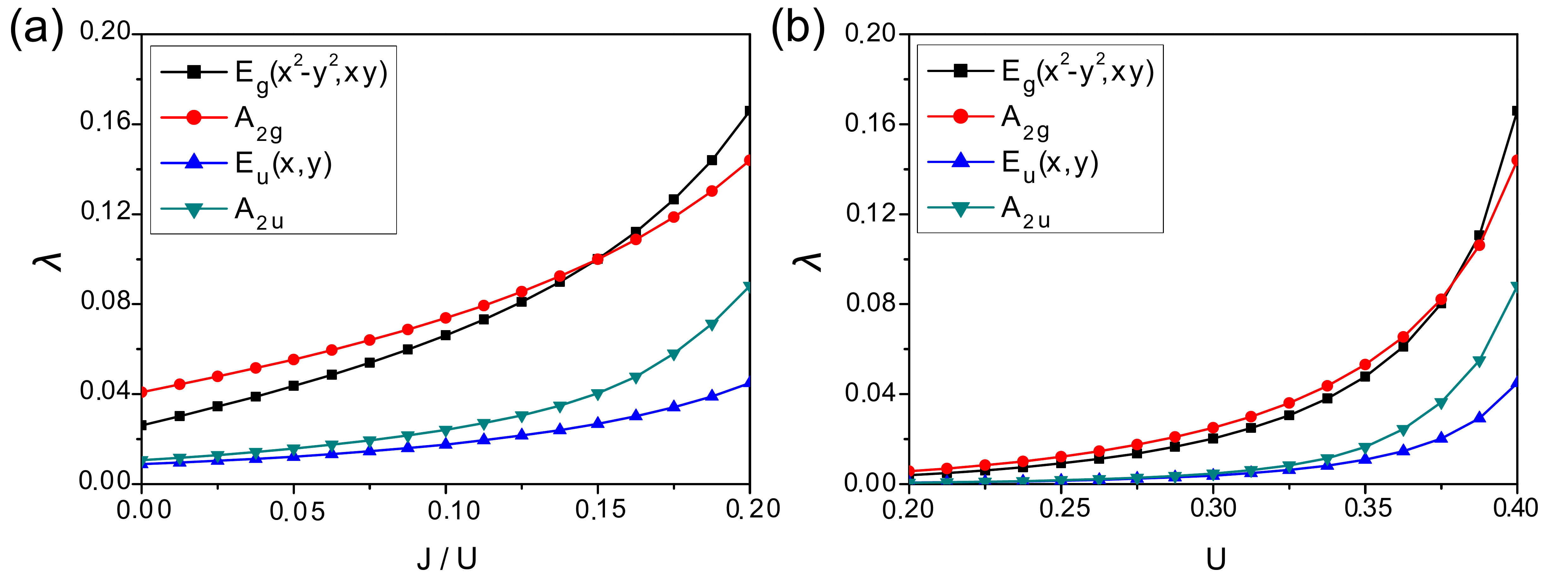}}
\caption{(color online) The two leading pairing strengths in singlet and triplet channels for superconducting state with 0.2 hole doping. The pairing strengths as a function of Hund's rule coupling parameter J/U with U=0.4 at 0.2 hole doping are plotted in (a). The pairing strengths as a function of on-site intraorbital coulomb interaction U with J/U=0.2 at 0.2 hole doping are plotted in (b).}
\label{lamda_hd}
\end{figure}

\begin{figure}
\centerline{\includegraphics[width=0.5\textwidth]{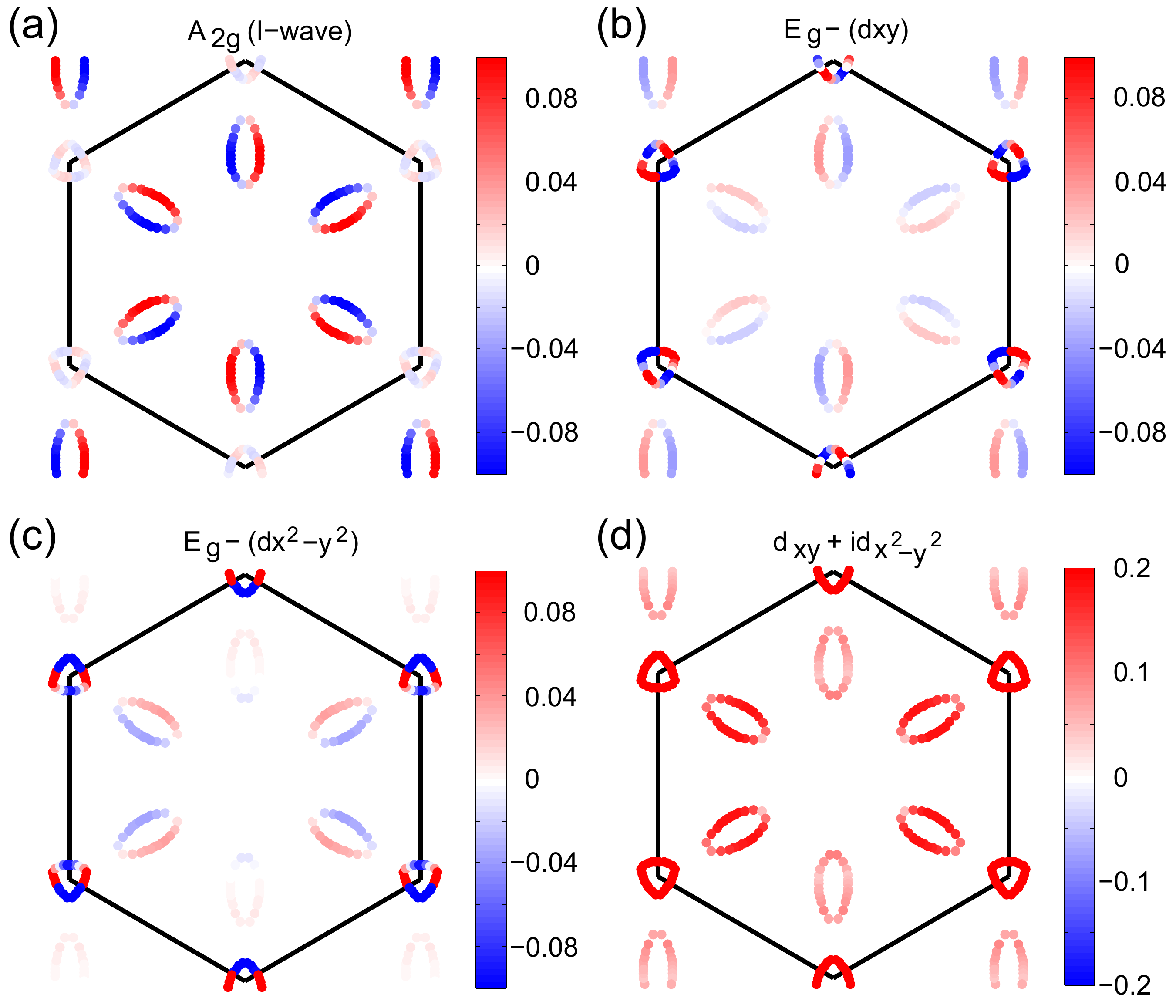}}
\caption{(color online) The gap functions of two leading singlet pairings with U=0.4 and J/U=0.2 at 0.2 hole doping. (a)  gap function of $A_{2g}$ (I-wave) irreducible representation. (b) and (c) are the gap functions of $d_{xy}$ and $d_{x^{2}-y^{2}}$ wave in $E_{g}$ irreducible representation. (d) gap function of $d_{xy}$+$id_{x^{2}-y^{2}}$.}
\label{pair_occ38}
\end{figure}

 Now we discuss the hole doped case with $\delta=-0.2$. Similar to the electron doped case, we find that $A_{2g}$ and $E_g$ pairing states are dominant and their pairing strengths are close, as shown in Fig. \ref{lamda_hd}. The A$_{2g}$ state is the leading pairing for small $J/U$($J/U < 0.15$) and the E$_{g}$ state becomes dominant pairing for large $J/U$ ($J/U > 0.15$). The corresponding gap functions of A$_{2g}$, $d_{x^{2}-y^{2}}$ and $d_{xy}$ are presented in Fig. \ref{pair_occ38}. The $A_{2g}$ state is very similar to the case with $\delta=0.2$, where gap functions almost vanish on $\beta$ FS. For $E_g$ state, the noticeable feature is the great enhancement of gap functions on $\beta$ compared with electron doped cases. Therefore $A_{2g}$ and $E_g$ pairing state are quite robust in doped monolayer NiPX$_{3}$.

%We fit the $A_{2g}$ and $E_g$ gap functions with lattice harmonics and investigate their real space structure.

 To further understand the obtained pairing state, we analyze the real-space structure of the obtained pairing states. In multiorbital system, the pairing in orbital space can also transform nontrivially under point group operations. For this two-band honeycomb lattice model, the classification of pairing states in real space have been given in Ref.\cite{WuTBG}. Generally the pairing state with $\eta$ form factor on $n$-th NN bond can be written as,
\begin{eqnarray}
\hat{\Delta}^{\eta}_n(\bm{k})=\Psi^\dag_{\bm{k}}F^{\eta}_n(\bm{k})[\Psi^\dag_{-\bm{k}}]^T=\Psi^\dag_{\bm{k}}[f^{\eta}_n(\bm{k})s_{i}\otimes \sigma_j \otimes \tau_l] [\Psi^\dag_{-\bm{k}}]^T,
\end{eqnarray}
where $\Psi^\dag_{\bm{k}}=(\psi^\dag_{\bm{k}\uparrow},\psi^\dag_{\bm{k}\downarrow})$ and $i,j,k=1,2,3$, $f^{\eta}_n(\bm{k})$ is the corresponding lattice harmonics in $\bm{k}$ space and $\bm{s}$, $\bm{\sigma}$, $\bm{\tau}$ are Pauli matrices defined in the spin, sublattice and orbital space. For the $A_{2g}$ state, the pairing matrix can be written as,
\begin{eqnarray}
F^{A_{2g}}(\bm{k})&=&\sum_{n\in AB}\Delta^{A_{2g}}_n[is_{2}\otimes (if^{E_1}_{nR} \sigma_1-if^{E_1}_{nI}\sigma_2  )\otimes \tau_3\nonumber\\
&&+is_{2}\otimes (f^{E_2}_{nR} \sigma_1-f^{E_2}_{nI}\sigma_2  )\otimes \tau_1]\nonumber\\
&&+\sum_{n\in AA/BB}\Delta^{A_{2g}}_n[is_{2}\otimes f^{E_2}_{n}\sigma_0\otimes \tau_3\nonumber\\
&&-is_{2}\otimes f^{E_1}_{n}\sigma_0\otimes \tau_1].
\end{eqnarray}
Here $f^{\eta}_{nI/R}$ represents the real/imaginary part of $\eta$ form factor for $n$-th NN bond. Because there is no $A_{2g}$ lattice harmonic, $A_{2g}$ pairing state must involve two-dimensional irreducible representations of lattice harmonics and orbital pairing.  For the $E_g$ state, the matrices for two pairing states can be written as,
\begin{eqnarray}
F^{E_{1g}}(\bm{k})&=&\sum_{n\in AB} \{\Delta^{A_{1g}}_nis_{2}\otimes (f^{A_{1}}_{nR}\sigma_1-f^{A_{1}}_{nI}\sigma_2)\otimes \tau_3\nonumber\\
&&+\Delta^{E}_n[is_{2}\otimes (f^{E_2}_{nR} \sigma_1-f^{E_2}_{nI}\sigma_2  )\otimes \tau_3\nonumber\\
&&+is_{2}\otimes (if^{E_1}_{nR} \sigma_1-if^{E_1}_{nI}\sigma_2  )\otimes \tau_1]\}\nonumber\\
&&+\sum_{n\in AA/BB}[\Delta^{A_{1g}}_nis_{2}\otimes f^{A_{1g}}_{n}\sigma_0\otimes \tau_3\nonumber\\
&&+\Delta^{E}_n(is_{2}\otimes f^{E_1}_{n}\sigma_0\otimes \tau_3- is_{2}\otimes f^{E_2}_{n}\sigma_0\otimes \tau_1) ],\nonumber\\
F^{E_{2g}}(\bm{k})&=&\sum_{n\in AB} \{\Delta^{A_{1g}}_nis_{2}\otimes (f^{A_{1g}}_{nR}\sigma_1-f^{A_{1g}}_{nI}\sigma_2)\otimes \tau_1\nonumber\\
&&+\Delta^{E}_n[is_{2}\otimes (if^{E_1}_{nR} \sigma_1-if^{E_1}_{nI}\sigma_2  )\otimes \tau_3\nonumber\\
&&+is_{2}\otimes (-f^{E_2}_{nR} \sigma_1+f^{E_2}_{nI}\sigma_2  )\otimes \tau_1]\}\nonumber\\
&&+\sum_{n\in AA/BB}[\Delta^{A_{1}}_nis_{2}\otimes f^{A_{1}}_{n}\sigma_0\otimes \tau_1\nonumber\\
&&+\Delta^{E}_n(is_{2}\otimes f^{E_2}_{n}\sigma_0\otimes \tau_3+is_{2}\otimes f^{E_1}_{n}\sigma_0\otimes \tau_1) ],\nonumber\\
\end{eqnarray}
The lattice harmonics are listed in the Appendix. We fit the obtained gap function from RPA calculations with the above form factors up to the third NN bonds.  The $I$-wave state is dominantly contributed by pairing on the NN and NNN bonds. In contrast, $E_g$ state in real space is extended and pairing on these bonds as well as onsite pairing will contribute.

\begin{figure}
\centerline{\includegraphics[width=0.5\textwidth]{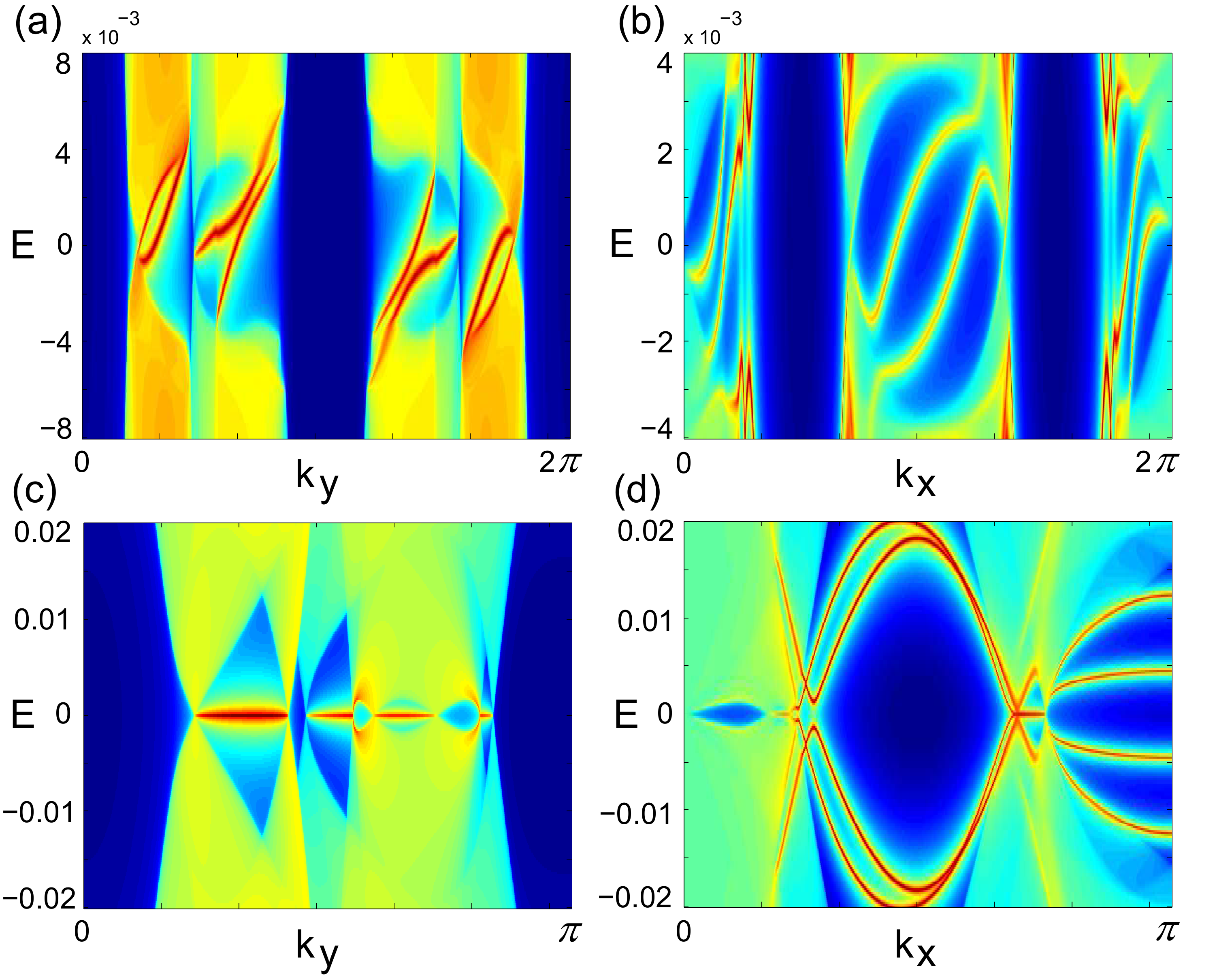}}
\caption{(color online) Zigzag and armchair edge spectra for the chiral $d$ wave (top panel) and $I$ wave pairing states (bottom panel) at 0.2 electron doping. Onsite $d$-wave pairing and NN $I$-wave pairing are adopted in the calculations.  }
\label{edge}
\end{figure}

Both of the obtained $d$-wave and $I$-wave pairing states can carry topological characters. The chiral $d$-wave state, breaking the time reversal symmetry, belongs to the class C and is characterized by the topological invariant $2Z$. For the nodal superconducting state can be characterized by a nonzero one-dimensional topological invariant\cite{Sato2011}, which can result dispersionless Andreev bound states (ABS) on surfaces or edges. For the chiral $d$-wave state, we plot the armchair and zigzag edges states with only including onsite $d$-wave state in Fig.\ref{edge} (a) and (b). As there are four Fermi surfaces in half Brillouin zone, the total Chern number is $4\times2=8$ per spin channel. In the both case, there are eight edge states, which is consistent with Chern number. With furthering including Rashba spin-orbit coupling and Zeeman coupling, odd Chern number can be achieved, which results edge Majorana modes\cite{Lu2018}. For the $I$-wave state, Fig. \ref{edge}(c) and (d) show the zigzag and armchair edge states. There are flat ABS connecting the projections of nodal points and their appearance is characterized by the 1D topological invariant, which is provided in the Appendix. One prominent feature of $I$-wave state is that flat ABS at zero energy can appear at all lattice termination edges, in sharp contrast to the $d$-wave pairing in cuprates. This will induce a large peak at zero energy in the density of states, which can be detected in STM measurements.

\section{ Discussion and Conclusion}\label{sectionv}

In order to achieve superconductivity in NiPX$_3$(X=S,Se), charge doping is required to suppress the antiferromagnetic order. For 2D layered materials, electron or hole carriers can be introduced by gating technology, which has been used to realize superconductivity for semiconducting thin films\cite{Ye2012}. Similar to graphene and monolayer FeSe, carries doping can be introduced by the adsorption of cations on the NiPX$_3$ films or cations intercalation in bulk NiPX$_3$. The charge doping in experiments can be easily realized in NiPX$_3$.

Once superconductivity is realized in Ni-based transition-metal trichalcogenides, $A_{2g}$ ($I$-wave) and $E_g$ ($d+id$) pairing states are dominant according to our calculations. $I$-wave state is characterized by its line nodes along $\Gamma-K$. While the chiral $d$-wave state, which has a full gap and breaks the time reversal symmetry, is characterized by a nontrivial topological invariant. For $I$-wave state, the nodal gap structure can be directly detected by the high resolution ARPES. Physical properties related to low energy excitations,
such as low temperature specific heat, spin relaxation and penetration depth, should be very similar to the $d$-wave state in cuprates. The time reversal symmetry breaking for the fully gapped $d+id$ state can be verified by muon-spin-rotation/relaxation measurements. Furthermore, at the edges there is a large peak at zero energy in DOS for $I$-wave state, which makes it distinct from the chiral $d$-wave state.  Therefore, experimental measurements in the superconducting states can be used to distinguish these two pairing states.

In summary, we have investigated the pairing symmetry for the Ni-based transition-metal trichalcogenide proposed recently. By performing RPA calculations, we find that $I$-wave ($A_{2g}$) state and chiral $d$-wave ($E_g$) state are dominant and nearly degenerate for typical electron and hole doping. Both of them are promoted by the intra Fermi surface nesting (between $\alpha$ FS) and the inter Fermi surface nesting (between $\alpha$ and $\beta$ FS). Their nontrivial topological properties are manifested by the presence of edge states. Ni-based transition-metal trichalcogenide provides us a new platform to study the exotic phenomena emerged from the electron-electron correlation.

\section{Acknowledgements}
This work is supported by the Ministry of Science and Technology of China 973 program (Grant No. 2015CB921300, No. 2017YFA0303100, No.2017YFA0302900), National Science Foundation of China (Grant No. NSFC-11334012), and the Strategic Priority Research Program of CAS (Grant No. XDB07000000). The work in W\"urzburg is funded by the Deutsche Forschungsgemeinschaft (DFG, German Research Foundation)-Project-ID 258499086-SFB 1170 and by the W\"urzburg-Dresden Cluster of Excellence on Complexity and Topology in Quantum Matter -- ct.qmat (EXC 2147, Project-ID 39085490).

\clearpage
\begin{widetext}

\appendix

\section{The tight-binding model of NiPS$_{3}$}
The four band tight-binding model is given by
\begin{equation}
\emph{H}_{0}=\sum_{\alpha\beta}\sum_{\mu\nu\sigma}h^{\alpha\beta}_{\mu\nu}(k)c^{\dag}_{\alpha\mu\sigma}c_{\beta\nu\sigma},
\end{equation}
where ($\alpha$,$\beta$) are the sublattice indices (a,b) and ($\mu$,$\nu$) are the orbital indices ($d_{xz}$, $d_{yz}$). $c^{\dag}_{\alpha\mu\sigma}$ creates a spin $\sigma$ electron in $\mu$ orbital on $\alpha$ sublattice. The matrix elements of $h^{\alpha\beta}_{\mu\nu}(k)$ based on the basis $c^{\dag}_{\alpha\mu}=(c^{\dag}_{A,xz},c^{\dag}_{A,yz},c^{\dag}_{B,xz},c^{\dag}_{B,yz})$ are listed in the follow:
\begin{equation}
\begin{aligned}
\emph{h}_{33}=&\emph{h}_{11}=(t_{21}+3t_{23})cos(\frac{\sqrt{3}k_x}{2})cos(\frac{k_y}{2})+2t_{21}cos(k_y)-\lambda,\\
\emph{h}_{44}=&\emph{h}_{22}=(3t_{21}+t_{23})cos(\frac{\sqrt{3}k_x}{2})cos(\frac{k_y}{2})+2t_{23}cos(k_y)-\lambda,\\
\emph{h}^{*}_{21}=&\emph{h}_{12}=[-4it_{22}cos(\frac{\sqrt{3}k_x}{2})+4it_{22}cos(\frac{k_y}{2})+\sqrt{3}(t_{21}-t_{23})sin(\frac{\sqrt{3}k_x}{2})]sin(\frac{k_y}{2}),\\
\emph{h}^{*}_{31}=&\emph{h}_{13}=\frac{1}{2}[cos(\frac{k_x}{2\sqrt{3}})+isin(\frac{k_x}{2\sqrt{3}})][(t_{11}+3t_{12})cos(\frac{k_y}{2})\\
&+cos(\frac{\sqrt{3}k_x}{2})(2(t_{11}+t_{31})+(t_{31}+3t_{32})cos(k_y))-i(2(t_{11}-t_{31})+(t_{31}+3t_{32})cos(k_y))sin(\frac{\sqrt{3}k_x}{2})],\\
\emph{h}^{*}_{41}=&\emph{h}_{14}=i\frac{3}{2}[(t_{11}-t_{12})cos(\frac{k_x}{2\sqrt{3}})sin(\frac{k_y}{2})+i(t_{11}-t_{12})sin(\frac{k_x}{2\sqrt{3}})sin(\frac{k_y}{2})\\
&-(t_{31}-t_{32})cos(\frac{k_x}{2\sqrt{3}})^2sin(k_y)+(t_{31}-t_{32})sin(\frac{k_x}{2\sqrt{3}})^2sin(k_y)+i(t_{31}-t_{32})sin(\frac{k_x}{\sqrt{3}})sin(k_y)],\\
\emph{h}^{*}_{32}=&\emph{h}_{23}=\emph{H}_{14},\\
\emph{h}^{*}_{42}=&\emph{h}_{24}=\frac{1}{2}[cos(\frac{k_x}{2\sqrt{3}})+isin(\frac{k_x}{2\sqrt{3}})][(3t_{11}+t_{12})cos(\frac{k_y}{2})\\
&+cos(\frac{\sqrt{3}k_x}{2})(2(t_{12}+t_{32})+(3t_{31}+t_{32})cos(k_y))-i(2(t_{12}-t_{32})+(3t_{31}+t_{32})cos(k_y))sin(\frac{\sqrt{3}k_x}{2})],\\
\emph{h}^{*}_{43}=&\emph{h}_{34}=[4it_{22}cos(\frac{\sqrt{3}k_x}{2})-4it_{22}cos(\frac{k_y}{2})+\sqrt{3}(t_{21}-t_{23})sin(\frac{\sqrt{3}k_x}{2})]sin(\frac{k_y}{2}).\\
\end{aligned}
\end{equation}
Here, (1,2,3,4) are orbital indices of ($d^{A}_{xz},d^{A}_{yz},d^{B}_{xz},d^{B}_{yz}$) and $\lambda$ is the chemical potential.

The hopping parameters in the model are
\begin{equation}
\begin{aligned}
&t_{11} = -0.036294,t_{12} = -0.050971,\\
&t_{21} = -0.015141,t_{22} = 0.003175,t_{23} = 0.012118,\\
&t_{31} = 0.238574,t_{32} = -0.020218.
\end{aligned}
\end{equation}

\section{lattice harmonics in honeycomb lattice}
The form factor for onsite pairing is $f^{A_1}_{0}=1$. The form factors for nearest-neighbor (NN) bond is,
\begin{eqnarray}
A_1 &&\quad f_1^{A_1}(k)=e^{\frac{ik_x}{\sqrt{3}}}+2e^{-\frac{ik_x}{2\sqrt{3}}}cosk_y/2, \\
E_2 &&\quad f_1^{E_2}(k)=e^{\frac{ik_x}{\sqrt{3}}}-e^{-\frac{ik_x}{2\sqrt{3}}}cosk_y/2 \sim \frac{\sqrt{3}}{2}ik_x, \\
E_1 &&\quad f_1^{E_1}(k)=\sqrt{3}e^{-\frac{ik_x}{2\sqrt{3}}}sink_y/2 \sim \frac{\sqrt{3}}{2}k_y.
\end{eqnarray}
The form factors for next NN bond is,
\begin{eqnarray}
A_1 &&\quad f_2^{A_1}(k)=2cosk_y+4cos\frac{\sqrt{3}}{2}k_xcos\frac{1}{2}k_y, \\
E_1 &&\quad f_2^{E_1}(k)=2cosk_y-2cos\frac{\sqrt{3}}{2}k_xcos\frac{1}{2}k_y\sim \frac{3}{4}(k^2_x-k^2_y),\\
E_2 &&\quad f_2^{E_2}(k)=2\sqrt{3}sin\frac{\sqrt{3}}{2}k_xsin\frac{1}{2}k_y \sim \frac{3}{4}2k_xk_y.
\end{eqnarray}
The form factors for third NN bond is,
\begin{eqnarray}
A_1 &&\quad f_3^{A_1}(k)=e^{-\frac{2ik_x}{\sqrt{3}}}+2e^{\frac{2ik_x}{\sqrt{3}}}cosk_y, \\
E_2 &&\quad f_3^{E_2}(k)=e^{-\frac{2ik_x}{\sqrt{3}}}-e^{\frac{ik_x}{\sqrt{3}}}cosk_y \sim {\sqrt{3}}ik_x, \\
E_1 &&\quad f_3^{E_1}(k)=\sqrt{3}e^{\frac{ik_x}{\sqrt{3}}}sink_y \sim {\sqrt{3}}k_y.
\end{eqnarray}

\section{topological number in $I$-wave state}
For time-reversal-invariant superconductor, the topological criterion\cite{Sato2011} about the zero energy ABS can be written as the following simple summation:
\begin{equation}
\omega(k_{y})=\frac{1}{2}\sum_{\varepsilon(k)=0}sgn[\partial_{k_{x}}\varepsilon(k)]\cdot sgn[\Delta(k)],
\end{equation}
where the summation is taken for $k_{x}$ satisfying $\varepsilon(k)=0$ with a fixed $k_{y}$.
According to the above formula, we obtain $\omega(k_{y})$ and $\omega(k_{x})$ about ABS of $I$-wave state in Tables \ref{ky} and \ref{kx}.
These topological numbers $\omega(k_{y})$ and $\omega(k_{x})$ are consistent with the zigzag and armchair edge of $I$-wave superconducting state respectively.

\begin{table}
\addtolength{\tabcolsep}{15pt}
\caption{Topological number $\omega(k_{y})$ for $I$-wave state with $\Delta=0.005$ eV. Here, $k^{i}_{y}$ ($i=1,2,3,4,5,6,7,8$)
are defined in Fig. \ref{appendix01}.}

\renewcommand{\arraystretch}{1.3}
\begin{tabular}{c|c}
\hline
\hline
%\multicolumn{5}{|c|}{\delta=0.88}
&$\omega(k_{y})$    \\
\hline
 $k^{7}_{y}<k_{y}<k^{8}_{y}$   &  1 \\
%\hline
 $k^{6}_{y}<k_{y}<k^{7}_{y}$   &  0 \\
%\hline
 $k^{5}_{y}<k_{y}<k^{6}_{y}$   & -2 \\
%\hline
 $k^{4}_{y}<k_{y}<k^{5}_{y}$   &  0 \\
%\hline
 $k^{3}_{y}<k_{y}<k^{4}_{y}$   & -1 \\
%\hline
 $k^{2}_{y}<k_{y}<k^{3}_{y}$   &  0 \\
%\hline
 $k^{1}_{y}<k_{y}<k^{2}_{y}$   & -2 \\
  \hline
  \hline
\end{tabular}
\label{ky}
\end{table}

\begin{table}
\addtolength{\tabcolsep}{15pt}
\caption{Topological number $\omega(k_{x})$ for $I$-wave state with $\Delta=0.005$ eV. Here, $k^{i}_{x}$ ($i=1,2,3,4,5$)
are defined in Fig. \ref{appendix01}.}
\renewcommand{\arraystretch}{1.3}
\begin{tabular}{c|c}
\hline
\hline
%\multicolumn{5}{|c|}{\delta=0.88}
&$\omega(k_{x})$    \\
\hline
 $k^{4}_{x}<k_{x}<k^{5}_{x}$   &  -2 \\
%\hline
 $k^{3}_{x}<k_{x}<k^{4}_{x}$   &  0 \\
%\hline
 $k^{2}_{x}<k_{x}<k^{3}_{x}$   &  2 \\
%\hline
 $k^{1}_{x}<k_{x}<k^{2}_{x}$   &  0 \\
  \hline
  \hline
\end{tabular}
\label{kx}
\end{table}

\begin{figure}
\centerline{\includegraphics[width=0.5\textwidth]{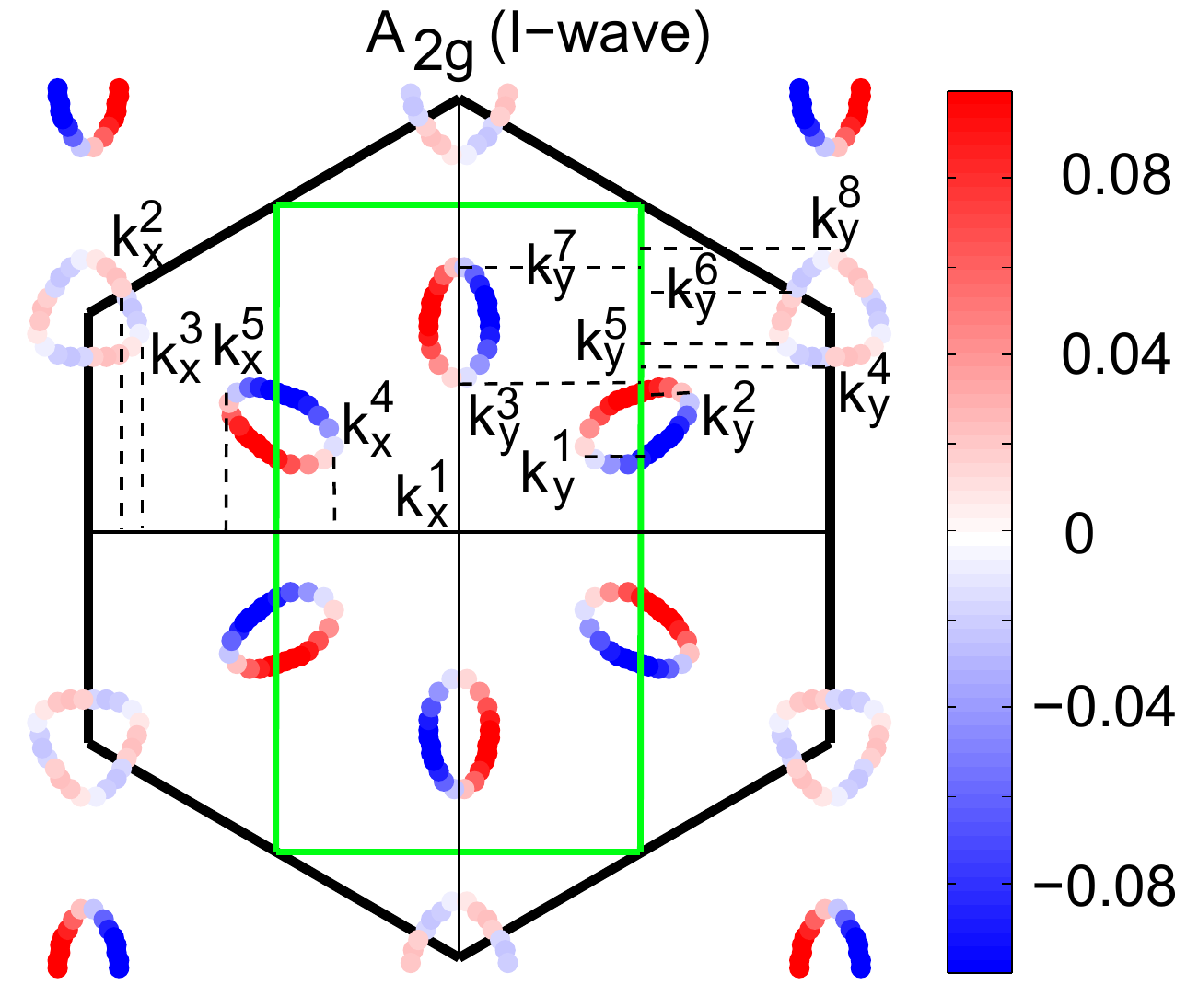}}
\caption{(color online) The gap function of $A_{2g}$ (I-wave) irreducible representation with U=0.4 and J/U=0.2 at 0.2 electron doping. The green line is the folded Brillouin zone for Zigzag edge state in momentum space.}
\label{appendix01}
\end{figure}

\end{widetext}

%\bibliographystyle{prsty}
%\bibliography{biblio}

\begin{thebibliography}{Iwave}
\bibitem{Nov} K. S. Novoselov, A. K. Geim, S. V. Morozov, D. Jiang, Y. Zhang, S. V. Dubonos, I. V. Grigorieva, and A. A. Firsov, Science \textbf{306}, 666 (2004).
\bibitem{Geim} A. K. Geim and I. V. Grigorieva, Nature \textbf{499}, 419 (2013).
\bibitem{Xu2014} X. D. Xu, W. Yao, D. Xiao, and T. F. Heinz, Nat. Phys. \textbf{10}, 343 (2014).
\bibitem{Kim} S. Y. Kim, T. Y. Kim, L. J. Sandilands, S. Sinn, M. C. Lee, J. Son, S. Lee, K. Y. Choi, W. Kim, B. G. Park, C. Jeon, H. D. Kim, C. H. Park, J. G Park, S. J. Moon, and T. W. Noh, Phys. Rev. Lett. \textbf{120}, 136402 (2018).
\bibitem{Kim2} K. Kim, S. Y. Lim, J. U. Lee, S. Lee, T. Y. Kim, K. Park, G. S. Jeon, C. H. Park, J. G. Park, and H. Cheong, Nat. Commun. \textbf{10}, 345 (2019)
\bibitem{Wildes} A. R. Wildes, V. Simonet, E. Ressouche, G. J. McIntyre, M. Avdeev, E. Suard, S. A. J. Kimber, D. Lancon, G. Pepe, B. Moubaraki, and T. J. Hicks, Phys. Rev. B \textbf{92}, 224408 (2015).
\bibitem{Gong} C. Gong, L. Li, Z. Li, H. Ji, A. Stern, Y. Xia, T. Cao, W. Bao, C. Wang, Y. Wang, Z. Q. Qiu, R. J. Cava, S. G. Louie, J. Xia, and X. Zhang, Nature \textbf{546}, 265 (2017).
\bibitem{Huang} B. Huang, G. Clark, E. Navarro-Moratalla, D. R. Klein, R. Cheng, K. L. Seyler, D. Zhong, E. Schmidgall, M. A. McGuire, D. H. Cobden, W. Yao, D. Xiao, P. Jarillo-Herrero, and X. Xu,                      Nature \textbf{546}, 270 (2017).
\bibitem{Lu} J. Lu, O. Zheliuk, I. Leermakers, N. F. Yuan, U. Zeitler, K. T. Law, and J. Ye, Science \textbf{350}, 1353 (2015).
\bibitem{Xi} X. Xi, Z. Wang, W. Zhao, J.-H. Park, K. T. Law, H. Berger, L. Forr\'{o}, J. Shan, and K. F. Mak, Nat. Phys. \textbf{12}, 139 (2015).
\bibitem{Saito} Y. Saito, Y. Nakamura, M. S. Bahramy, Y. Kohama, J. Ye, Y. Kasahara, Y. Nakagawa, M. Onga, M. Tokunaga, T. Nojima, Y. Yanase, and Y. Iwasa , Nat. Phys. \textbf{12}, 144 (2016).
\bibitem{Costanzo} D. Costanzo, S. Jo, H. Berger, and A. F. Morpurgo, Nat. Nanotechnol, \textbf{11}, 339 (2016).
\bibitem{Wang} Y. G. Wang, J. J. Ying, Z. Y. Zhou, J. L. Sun, T. Wen, Y. N. Zhou, N. N. Li, Q. Zhang, F. Han, Y. M. Xiao, P. Chow, W. G. Yang, V. V. Struzhkin, Y. S. Zhao, and H. K. Mao,
Nat. Commun. \textbf{9}, 1914 (2018).
\bibitem{Chittari} B. L. Chittari, Y. Park, D. Lee, M. Han, A. H. Macdonald, E. Hwang, and J. Jung, Phys. Rev. B \textbf{94}, 184428 (2016).
\bibitem{Sivadas} N. Sivadas, M. W. Daniels, R. H. Swendsen, S. Okamoto, and D. Xiao, Phys. Rev. B \textbf{91}, 235425 (2015).
\bibitem{Lee} J. U. Lee, S. Lee, J. H. Ryoo, S. Kang, T. Y. Kim, P. Kim, C. H. Park, J. Park, and H. Cheong, Nano Lett. \textbf{16}, 7433 (2016).
\bibitem{Kuo} C. T. Kuo, M. Neumann, K. Balamurugan, H. J. Park, S. Kang, H. W. Shiu, J. H. Kang, B. H. Hong, M. Han, T. W. Noh, and J. G. Park, Sci. Rep. \textbf{6}, 20904 (2016).


\bibitem{YHGu} Y. H. Gu, Q. Zhang, C. C. Le, Y. X. Li, T. Xiang, and J. P. Hu, arXiv:1811.02333.
\bibitem{Flem} G. Le Flem, R. Brec, G. Ouvard, A. Louisy, and P. Segransan, J. Phys. Chem. Sol. \textbf{43}, 455 (1982).
\bibitem{Bickers} N. E. Bickers, D. J. Scalapino, and S. R. White, Phys. Rev. Lett. \textbf{62}, 961 (1989).
\bibitem{Kemper} A. F. Kemper, T. A. Maier, S. Graser, H. P. Cheng, P. J. Hirschfeld, and D. J. Scalapino, New J. Phys. \textbf{12}, 073030 (2010).
\bibitem{Graser} S. Graser, T. A. Maier, P. J. Hirschfeld, and D. J. Scalapino, New J. Phys. \textbf{11}, 025016 (2009).
\bibitem{Xxwu1} X. X. Wu, J. Yuan, Y. Liang, H. Fan, and J. P. Hu, Europhys. Lett. \textbf{108}, 27006 (2014).
\bibitem{Xxwu2} X. X. Wu, F. Yang, C. C. Le, H. Fan, and J. P. Hu, Phys. Rev. B \textbf{92}, 104511 (2015).
\bibitem{Li} Y. X. Li, X. L. Han, S. S. Qin, C. C. Le, Q. H. Wang, and J. P. Hu, Phys. Rev. B \textbf{96}, 024506 (2017).
\bibitem{Sante} D. D. Sante, X. X Wu, Mario Fink, Werner Hanke, and Ronny Thomale, arXiv:1809.03970.
\bibitem{Graphene3RG}
R. Nandkishore, L.~S. Levitov, and A.~V. Chubukov, Nat. Phys. {\bf 8},  158
  (2012).
\bibitem{GrapheneFRG}
M.~L. Kiesel, C. Platt, W. Hanke, D.~A. Abanin, and R. Thomale, Phys. Rev. B
  {\bf 86},  020507  (2012).
\bibitem{GrapheneFRG2}
W.-S. Wang, Y.-Y. Xiang, Q.-H. Wang, F. Wang, F. Yang, and D.-H. Lee, Phys.
  Rev. B {\bf 85},  035414  (2012).
\bibitem{CobaltatesFRG}
M.~L. Kiesel, C. Platt, W. Hanke, and R. Thomale, Phys. Rev. Lett. {\bf 111},
  097001  (2013).
\bibitem{PhysRevB.89.020509}
M.~H. Fischer, T. Neupert, C. Platt, A.~P. Schnyder, W. Hanke, J. Goryo, R.
  Thomale, and M. Sigrist, Phys. Rev. B {\bf 89},  020509  (2014).
\bibitem{Sato2011} M. Sato, Y. Tanaka, K. Yada, and T. Yokoyama, Phys. Rev. B \textbf{83}, 224511 (2011).
\bibitem{Lu2018} C. Lu, L. D. Zhang, X. X. Wu, F. Yang, and J. P. Hu Phys. Rev. B \textbf{97}, 165110 (2018).
\bibitem{WuTBG} X. Wu, {\em et al.}, in preparation (2019).
\bibitem{Ye2012} J. T. Ye, Y. J. Zhang, R. Akashi, M. S. Bahramy, R. Arita, and Y. Iwasa, Science \textbf{338}, 1193 (2012).




\end{thebibliography}

\end{document}